\documentstyle[12pt]{article}
\catcode`\@=11
\begin{document}
\def \sin {{\rm sin}}
\def \cos {{\rm cos}}
\def \tan {{\rm tan}}
\def \cot {{\rm cot}}

\def \A {{\cal A}}
\def \AA {\td {\cal A}}
\def \third {{\textstyle {1\ov 3} } }
\def\Jo#1#2#3#4{{#1} {\bf #2}, #3 (#4)}
\def \re#1{(\ref{#1})}
\def\st{\scriptstyle}
\def\sst{\scriptscriptstyle}
\def\mco{\multicolumn}
\def\epp{\epsilon^{\prime}}
\def\vep{\varepsilon}
\def\ra{\rightarrow}
\def\vp{{\bf p}}
\def\al{\alpha}
\def\ab{\bar{\alpha}}
\def \bi{\bibitem}
\def \ep{\epsilon}
\def\D{\Delta}
\def\sms{$\s$-models }
\def \om {\omega}
\def \foot{\footnote}
\def\be{\begin{equation}}
\def\ee{\end{equation}}
\def \lab {\label}
\def \k {\kappa} 
\def \F {{\cal F}}
\def \g {\gamma}
\def \del {\partial}
\def \bd {\bar \partial }
\def \na {\nabla}
\def \const {{\rm const}}
\def \ha{{\textstyle{1\over 2}}}
\def \na {\nabla }
\def \D {\Delta}
\def \a {\alpha}
\def \b {\beta}
\def \chi {\chi}\def\r {\rho}
\def \s {\sigma}
\def \p {\phi}
\def \m {\mu}
\def \n {\nu}
\def \vp {\varphi }
\def \l {\lambda}
\def \t {\theta}
\def \td {\tilde }
\def \d {\delta}
\def \ci {\cite}
\def \la {\label}
\def \sm {$\s$-model }
\def \foot {\footnote }
\def \P {\Phi}
\def \o {\omega}
\def \inv {^{-1}}
\def \ov {\over }
\def \four{{\textstyle{1\over 4}}}
\def \fourth{{{1\over 4}}}
\def \foot{\footnote}
\def\be{\begin{equation}}
\def\ee{\end{equation}}
\def\bea{\begin{eqnarray}}
\def\eea{\end{eqnarray}}
\def\np {{\em  Nucl. Phys. }}
\def \pl {{\em  Phys. Lett. }}
\def \mpl {{\em Mod. Phys. Lett. }}
\def \prl {{ \em  Phys. Rev. Lett. }}
\def \pr  {{\em  Phys. Rev. }}
\def \ap  {{\em Ann. Phys. }}
\def \cmp {{\em Commun.Math.Phys. }}
\def \ijmp {{\em Int. J. Mod. Phys. }}
\def \jmp {{\em J. Math. Phys.}}
\def \cqg {{\em  Class. Quant. Grav. }}

\begin{titlepage}
\begin{flushright}Imperial/TP/95-96/32\\hep-th/9603099\\
March 1996\\
\end{flushright}
\vskip 3cm
\begin{center}
{\Large\bf Generalised  chiral null models   }\\
\vskip 0.2cm
{\Large\bf  and rotating string backgrounds }
\vskip 1.5cm
{\bf A.A. Tseytlin\footnote{e-mail: tseytlin@ic.ac.uk. \  On leave from Lebedev Institute, Moscow.}}\\
\vskip 0.2cm
{\it Theoretical Physics Group, Blackett Laboratory }\\
{\it Imperial College,   London SW7 2BZ, U.K.}
\end{center}

\begin{abstract}
{We consider an extension of a special class of conformal sigma models 
(`chiral null models') which describe extreme supersymmetric string solutions.
The new models contain both `left' and `right' vector couplings
and should correspond to non-BPS backgrounds. In particular, we discuss a conformal six-dimensional model which is a combination of fundamental string 
and 5-brane models with the two extra couplings representing rotations
in the orthogonal planes. If the two rotation parameters are independent the  resulting background is found to be either singular or not asymptotically flat.
The non asymptotically flat solution has a regular short distance limit described by a `twisted' product of $SL(2,R)$ and $SU(2)$ WZW theories 
with two twist parameters mixing the isometric Euler angles of 
$SU(2)$ with a null direction of $SL(2,R)$.}
\end{abstract} 
\end{titlepage}

\section{Introduction}
Superstring solutions with vanishing R-R background fields are described by conformal $\s$-models \ci{call,T}.
Related solutions with non-vanishing R-R fields can be 
obtained, e.g., by applying $SL(2,Z)$ duality 
of $D=10$ type IIB superstring theory \ci{twob}. 
Knowing a \sm description of a given  NS-NS string solution is 
important since many of its  features can be (at least, qualitatively) understood using 
conformal field theory considerations 
(see, e.g., \ci{LW,USS,TT}). 
This  may complement the picture obtained 
for the corresponding R-R solution using D-brane approach
(see, e.g, \ci{SV,CAM,HS,MV}). 

An important class of exact 
(supersymmetric,  BPS saturated) string 
solutions  is described by a special $\s$-model
-- `chiral null model' (CNM)  \ci{TH}.  
Here the 
 equations on background fields 
 obtained   by imposing  the conformal invariance 
condition are 
decoupled Laplace-type equations. The solutions are expressed 
in terms of harmonic functions, i.e. can be freely superposed
(in agreement with the BPS nature of resulting  field configurations).
Remarkably, the `pure-electric' CNM  with flat transverse space 
\ci{TH} 
admits a `dyonic'  generalisation \ci{US,USS,TT}
where the  transverse space  is 
represented by a non-trivial $N=4$ supersymmetric \sm.
 
CNM is characterised by the presence of only certain special  chiral 
couplings in the 
action. It is of interest to study 
what happens when one introduces extra 
integrable marginal perturbations  getting as a result  a mixture 
of  `left' and `right'  couplings
(which, in general,  may  break supersymmetry).
The corresponding  conformal invariance conditions
will no longer factorize into a simple set of Laplace equations, 
i.e. one will no longer have a 
simple BPS-type superposition principle.
These  models  may  describe certain `non-BPS' 
solutions which may still have some special properties
(e.g.  may  still   be exact to all orders).
An example of such model will be  discussed below.
It presumably still corresponds to  extremal field configurations  
but with less (than in standard CNM case) or no supersymmetry.

Our investigation of such models was partly motivated 
by an attempt to construct  `rotational' generalisations
of $D=6$  CNM's  
describing  $D=4$ \ci{US,USS} and $D=5$ \ci{TT}  
extreme dyonic black holes (thus providing exact string-theory
generalisations of the solutions of  leading-order
effective equations in \ci{CY} and \ci{SV}). 
These $D=6$ conformal models have 
 a short-distance (horizon or `throat' \ci{chs,SL})  $r \to 0$ 
region described by 
(a factor of)
 $SL(2,R) \times SU(2)$   WZW  theory. The latter 
is remarkable in that  in the supersymmetric
case it  has  the free value
$c_{eff}=6$ of the central charge and thus the dilaton is constant
at the `throat'. 

Adding  one rotational parameter to a $D=5$ solution parametrised
 by 3 independent  charges (two electric and one `magnetic')
can be understood 
\ci{TT} as adding a perturbation 
($\del u \bar J_3$) which `mixes'   
the  Cartan $\bar J_3$ curent of $SU(2)$ with an isometric 
(Gauss decomposition) coordinate $u$  of $SL(2,R)$. 
It can be induced by shifting one of the isometric Euler angles of 
$SU(2)$, $\psi \to \psi + q_1 u$.\foot{Closely related 
`magnetic' models were discussed
in \ci{RU,KK,TSE}.} 
The  `integrated' CNM which extrapolates such  perturbed 
 throat region model to finite $r$ 
describes the 4-parameter (3 charges and two equal  angular momenta in two orthogonal planes) extreme $D=5$ black  
hole solution \ci{TT} which generalises
 the  solution of the leading-order effective string equations
originally  found
in \ci{MV} (where the 3 non-trivial charge parameters were all equal).

A natural idea is to generalise the throat region model 
further  shifting also the {\it second}
Euler angle of $SU(2)$, $\vp \to \vp + q_2 u$, 
thus inducing the perturbation $\bd u J_3$ 
which  should  correspond to  the second independent rotation 
parameter in the resulting $D=5$ background.
For conformal invariance one needs also to add non-trivial
$O(q_1q_2 \del u \bd u)$ term. Similar  string model  was considered 
   in \ci{TSE} in the context  of  
continuous  supersymmetry breaking by `magnetic' backgrounds. 

The  \sm which  extends such $(q_1,q_2)$-perturbed
$SL(2,R)\times SU(2)$ WZW model to   finite $r$ 
is  a  simple  generalisation of  CNM  discussed below. 
We shall find that for $q_2\not=0$ the conformal invariance conditions
do not have solutions which describe asymptotically 
flat at $r\to \infty$  {\it and} 
non-singular  at $r\to 0$  backgrounds.  
This implies that  there are no non-singular extremal $D=5$ black holes 
with two independent rotational parameters 
(in agreement with similar remarks in \ci{MV}). 

 The solution that reduces to the  
regular $(q_1,q_2)$-perturbed
 throat region model   is not asymptotically  flat 
and describes a rotating magnetic universe.
Analogous   solutions   were discussed in \ci{TH,TRU}.\foot{There is also a similarity  to Kaluza-Klein Melvin model 
\ci{KKME,RU}  which can be   obtained by shifting an angular coordinate by a Kaluza-Klein one.} 
Given that  the background with $q_1=q_2=0$ has a simple
`fundamental string + 5-brane' interpretation 
it may be of interest to study the 
D-brane description  of the direct analog of this 
$q_1q_2\not=0$ solution  which has 
 non-vanishing  R-R fields.
The $D$-brane  picture   in \ci{MV} based on two $U(1)$ currents
of $SU(2)\times SU(2)$ KM algebra of underlying 
2d conformal model  has a striking  similarity with
the above description of the throat region model
with  the two  (`left' and `right')
 Cartan current perturbations.

\section{Review}
The `standard' CNM with curved transverse part
is defined by the Lagrangian \ci{TH}\foot{We shall ignore possible $u$-dependence of couplings.
When  both $K$  and ${\cal A}_i$ 
depend on $u$ one can redefine $v$ to absorb $K$ into ${\cal A}_i$ 
\ci{TH}.} 
\be
   L =  F(x)  \del u \left[\bd v +
   K(x) \bd u  +   2{\cal A}_i(x)  \bd  x^i \right]
 +  
\ha {\cal R} \ln F(x) 
+ L_{\bot}\ , 
\la{laggg}
\ee
\be
L_{\bot}= (G_{ij} + B_{ij})(x) \del x^i \bd x^j    +    {\cal R}
\p (x)\  . 
 \la{lggg}
\ee
There exists a renormalisation  scheme  
in which (\ref{laggg}) is conformal to all 
orders in $\a'$ provided
  the `transverse'  \sm \re{lggg}  is conformal
   and  the functions 
$F\inv,K,{\cal A}_i, \P$ satisfy
certain conformal invariance conditions.
The simplest tractable case is when the transverse
theory is either flat or has at least $(4,0)$ 
extended  world-sheet supersymmetry so that the conformal invariance
conditions  essentially preserve their 1-loop form, i.e. 
are the `Laplace'  equations 
in the  `transverse' background 
  \ci{TH,USS}
\be
 \del_i   (e^{-2\p} \sqrt G G^{ij} \del_j F^{-1}) =0 \  ,  \ \ 
\ \ \ \del_i   (e^{-2\p} \sqrt G G^{ij} \del_j K) =0 \  ,
\la{LL}
\ee
\be
 \hat \nabla_{+ i }(e^{-2\p} {\cal F}^{ij} ) = 0\ , \ \ \ 
i.e. \ \  \   \del_i   (e^{-2\p} \sqrt G {\cal F}^{ij})  -
\ha e^{-2\p} {\sqrt G} H^{kij} {\cal F}_{ki}
=0  \  ,    
 \la{cond}
\ee
 where $\hat \Gamma^i_{jk}
=  \Gamma^i_{jk} + {1\over 2} H^i_{\ jk}  ,  \ \ 
 {\cal F}_{ij} \equiv  \partial_i {\cal A}_j -
\partial_j {\cal A}_i , \ H_{ijk}= 3\del_{[i}B_{jk]}.$
For example, in the case when 
the transverse space  is described by the `5-brane' model \ci{chs}
($i,j, ...=1,2,3,4$)
\be
L_{\bot}= f(x) \del x^i \bd x^i +   B_{ij }(x)  \del x^i \bd x^j  + 
{\cal R}
\phi (x)\  , 
\la{fiv}
 \ee
$$
    G_{ij} = f(x) \delta_{ij} \  ,  \ \ \ \ 
 H^{ijk} = -{ 2\ov \sqrt G}\epsilon^{ijkp} \del_p \p 
\ , \  \ \ \ \p= \ha \ln f\  , \ 
\ \ \del^i\del_i   f =0\ ,   
$$
the equations for $F,K,f$ become
the free 4-d Laplace equations 
$\del^i\del_i  F\inv= 0, \ \del^i\del_i  K= 0, $
while the equation for $\A_i$ 
can be re-written as follows  \ci{TT}
\be    
\del_i   (e^{-2\p}  \sqrt G  {\cal F}^{ij}_+ ) =0   \ ,
\ \ \ i.e.\ \  \ \ \del_i   (f\inv   {\cal F}^{ij}_+ ) =0 \ , 
\la{sell}
\ee
$$
 \F^{ij}_+\equiv  {\cal F}^{ij} + {\cal F^*}^{ij} \ , \ \ \ \ \ \ 
{\cal F^*}^{ij} = {1\ov 2 \sqrt G  }  \ep^{ijkl} {\cal F}_{kl}  \ .  $$
 This equation can be solved, e.g.,  by 
imposing 
${\cal F}^{ij}_+ =0$
 which is again a linear flat-space equation
(the scale factor of  conformally flat
$G_{ij}$  drops out)
and should  
  describe supersymmetric BPS-type backgrounds. A particular solution is the  4-parameter rotating dyonic $D=5$ black hole  \ci{TT} which generalises
the solution of \ci{MV}.

In general the above equation  for $\A_i$ 
depends on the  harmonic function $f$ of the transverse theory  and does not reduce to  a flat-space Maxwell equation.
This is illustrated by a related 
 example with an  extra 
`left' electric  charge  added to a dyonic model with two  electric and two magnetic functions  which 
was discussed in \ci{USS}. The resulting equation for 
the new electric charge function $A(x)$ (a component of $\A_i$)
was not a free flat-space Laplace equation.
Generically  there is an interaction between 
electric (longitudinal or time-like) and magnetic (transverse or space-like) parts of the model which cannot be switched off.\foot{This 
raises the following  
 question: 
 given  a type II or heterotic theory compactified to $D=4$,  are there 
`dyonic' solutions parametrised by 
more than  two  general (not necessarily one-center) 
 harmonic functions of 3 spatial coordinates?}

In addition to the `longitudinal'--`transverse'  coupling
one can  also  introduce 
an interaction  between `left-moving' and `right-moving'  coupling functions. The simplest example 
is provided by  the `plane-wave' model  \ci{TTT}
\be
   L =   \del u \bd v + 
   K(x)\del u  \bd u  +  
 2{\cal A}_i(x) \del u  \bd  x^i  +  
 2\bar {\cal A}_i(x)   \del   x^i \bd u + L_{\bot}\ , 
\la{leg}
\ee
or its $u$-dual \ci{TH} ($\td F=K\inv, \ \td \A_i  =\bar \A_i$,\  
 $v'= v +\td u$)
\be
L= \td F(x) [\del 
\td u - 2\td \A_i (x)\del x^i][\bd v' + 2\A_j(x) \bd x^j] + 
\ha {\cal R} \ln \td F(x) 
+ L_{\bot}\ .
\la{morr}
\ee
which are conformal provided
\be
\del_i   [e^{-2\p} \sqrt G {\cal F}^{ij}(\A) ]  -
\ha e^{-2\p} {\sqrt G} H^{kij} {\cal F}_{ki}(\A)
=0 \    ,    
\la{ccoo}
\ee
$$
 \ \  
\del_i   [e^{-2\p} \sqrt G {\cal F}^{ij}(\bar \A)]  + 
\ha e^{-2\p} {\sqrt G} H^{kij} {\cal F}_{ki}(\bar \A)
=0  \  , $$
$$
\del_i   (e^{-2\p} \sqrt G G^{ij} \del_j K) 
  -  2 e^{-2\p} \sqrt G \F^{ij}(\A)\F_{ij}(\bar \A) =0\  .
$$
The equation for $K$ (or $\td F\inv$) is  no longer 
a free Laplace equation, i.e. there is no  simple superposition principle. In contrast to the standard plane-wave case \ci{PW}
one may  also expect  higher $\a'$-corrections 
(unless there exists a  special `exact' scheme).
The special cases in which the simplicity is restored 
correspond to taking   the two vector field strengths to be  constant
or  choosing an anti-self-dual 
$\F_{ij}(\A)$ and self-dual $\F_{ij}(\bar \A)$
so that $\F^{ij}(\A)\F_{ij}(\bar \A)=0$, i.e.  the equation 
for  $K$ becomes again the  homogeneous Laplace one.

\section{Generalised CNM}
Below we shall  consider the following generalisation 
of the above models \re{laggg},\re{leg} 
\be
   L =  F(x)[ \del u \bd v + 
   K(x)\del u  \bd u  +  
 2{\cal A}_i(x) \del u  \bd  x^i  +  
 2\bar {\cal A}_i(x) \del  x^i\bd u  ]  + \ha {\cal R} \ln  F(x) + L_{\bot}\ . 
\la{lrg}
\ee
The  $u$-dual of this model is a generalisation of \re{morr}
\be
\td L= \td F(x) [\del 
\td u -  2\td \A_i (x)\del x^i][\bd v + \td K(x) \bd \td u + 
2\A_j(x) \bd x^j] + 
\ha {\cal R} \ln \td F(x) 
+ L_{\bot}\ ,  
\la{mrr}
\ee
\be
 \td F = K\inv, \ \ \ \ \td K = F\inv, \ \ \ \ 
\td \A_i =  F \bar \A_i \ . \la{ooo} 
\ee
In contrast to the original CNM \re{laggg} the 
generalised model \re{lrg}
is  no longer  self-dual. This is also an indication of  its `non-BPS' 
 nature.

The 1-loop conditions of conformal invariance of  \re{lrg}
can be derived   from the 
standard 
equation $\hat R_{-\m\n} + 2 \hat \nabla_{-\m} \hat \nabla_{-\n}\Phi= 0 $
 for a general \sm ($x^\m=(u,v,x^i))$. 
A more  illuminating way of obtaining them is to 
 add source terms 
($\Delta L= - \del u
\bd V - \bd v \del U$) and   to   integrate out $u$ and $v$ as in \ci{TH}.\foot{The fact that the action is still linear in $v$ 
so that the  null coordinates   $u$ and $v$ 
can be integrated out strongly indicates  
that there still exist a scheme in which 
the 1-loop  conformal invariance conditions are exact 
to all orders (provided  this is also 
true for the transverse part of the model).}
As a result,  we get
\be
L'= K\del U\bd \del\inv (F\inv \del U) - F\inv \del U \bd V 
\la{sou}
\ee
$$ + \ 2   \A_i \bd x^i \del U
 + 2 F \bar\A_i \del x^i \bd \del\inv (F\inv \del U) + L_{\bot}
\ . $$
The  $O(\ln F)$ dilaton term   cancels out.  Since $U$ and $V$ are external fields
it is then  straightforward to study conditions of conformal invariance
with respect to the transverse coordinates only.
One finds the following set of equations which generalises  \re{LL},\re{cond},\re{ccoo}
\be
 \del_i   (\sqrt G e^{-2\p}G^{ij} \del_j F^{-1}) =0 \  ,
\la{LLL}
\ee
\be
 \del_i   ( \sqrt G e^{-2\p}G^{ij} \del_j K) 
- 2  \sqrt G e^{-2\p} {\cal F}^{ij}(\AA)[ {\cal F}_{ij}(C)  - F\inv {\cal F}_{ij}(\AA)]
=0 \  ,
\la{LLLL}
\ee
\be
 \del_i   [\sqrt G e^{-2\p}  {\cal F}^{ij}(\td \A)]  + 
\ha  {\sqrt G e^{-2\p}  } H^{kij} {\cal F}_{ki}(\td \A)
=0  \  ,    
 \la{condw}
\ee
\be
 \del_i   \{ \sqrt G e^{-2\p} [ {\cal F}^{ij}(C) 
- 2 F\inv {\cal F}^{ij}(\td \A)] \}  - 
\ha  { \sqrt G e^{-2\p}  }  H^{kij} {\cal F}_{ki}(C) 
=0  \   .    
 \la{cdw}
\ee
Here the gauge-invariant  field strengths  
correspond to the following potentials\foot{As it is clear from 
\re{mrr} the dual action is invariant under 
$ \td u \to \td u + h(x), $  $ \  v\to  v + s(x), $ $
\A_m\to  \A_m - \del_m s - \td K \del_m h, $ $ \ 
\AA_m \to \AA_m +  \del_m h, $
so that 
 the gauge-invariant vector field strengths 
are $ d \AA$ and $ d\A+ dF\inv \wedge \AA= 
dC - F\inv d\AA$ or just $d\AA$ and $dC$.}
\be \td \A_i \equiv  F \bar \A_i\ , \ \ \ \ \ \  \ 
C_i \equiv  \A_i + \bar \A_i \ . 
\la{ac}
\ee
\subsection{Flat transverse space: 
charged fundamental string }
Let us first  discuss  an example of  generalised CNM \re{lrg}
with 
flat transverse 
space  but $\bar \A_i\not=0$ which corresponds to 
a generalisation of the charged 
$D=5$ fundamental string solution \ci{FSS}.
Splitting the coordinates  $x^i$ into  $N=5$ compact  toroidal 
ones $y^n$ and 3 non-compact spatial ones $x^s$
we may set  
\be
 \A= \A_n(x) dy^n\ , \ \ \ \   \bar \A= \bar \A_n(x) dy^n\ , 
\ \ \   \ \td \A_n = F\bar \A_n\ , \ \ \ \  
C_n = \A_n + \bar \A_n\ , 
\la{ppp}
\ee
 and assume
that  all the functions  will  depend only on $x^s$. 
Then the  equations \re{LLL}--\re{cdw}
become 
\be
 \del^s\del_s F^{-1} =0 \  , \ \ \ \ 
  \del^s\del_s K 
- 4  \del^s \td \A_n ( \del_s C_n 
  - F\inv \del_s \td \A_n)
=0 \  ,
\la{LoLL}
\ee
\be
 \del^s   (\del_s C_n -  2F\inv   \del^s \td \A_n )=0  \  ,  
  \ \ \ \ 
 \del^s  \del_s \AA_n =0 \  .    
 \la{ucdw}
\ee
Though for  $\bar \A_n\not=0$ these equations  no longer
reduce to  free Laplace equations,  their
general solution  can still be expressed 
in terms of $2N+2=12$ 
 independent 
harmonic functions.\foot{In the heterotic string case one may introduce extra 16  `right'  vector couplings  leading to extra 16 harmonic functions.} 
 Choosing  all harmonic functions  to be 1-center ones
we find  
\be 
F\inv=1 + Qr^{-1}  , \ \ \ 
\AA_n = p_n r^{-1}  ,  \ \ \
C_n= (p_n +q_n)r^{-1} + p_n Q r^{-2} , \la{uuu}
\ee
$$
  \A_n =  q_n r^{-1}\  ,  \ \ \  
\ \ \  \bar \A_n =  p_n r^{-1} + p_n Q r^{-2} \  , 
$$
\be
K=  1 + \td Q r^{-1}   + 2q_n p_n r^{-2} 
+ {\textstyle  {2\ov 3}} p_n p_n Qr^{-3} \ , \ \ \ \ r^2=x^sx^s\ .  
\la{yyy}
\ee
When  both $q_n$ and $p_n$ are non-vanishing 
the  off-diagonal components of the 
$D=10$ metric 
and the antisymmetric tensor are different, i.e. one finds  the fundamental 
string background with  both  `left' and `right' charges.
Separating the  `internal' $y^n$-part  
($(\del y_n + F C_n \del u)^2 + ...$)
the rest of the action becomes the  neutral 
fundamental string  CNM  with redefined function $K \to 
K'=K - F C_n^2$. 
The reduction to $D=4$ then leads to a family
of  electrically charged  black holes 
parametrised by the  charges $Q,\td Q, p_n, q_n$
(12 in type II  or 12 +16 in  the heterotic string case).
The corresponding $D=4$ Einstein-frame metric is given by 
\be
ds^2_E = - \l (r) dt^2 + \l^{-1}(r) (dr^2 + r^2 d\Omega)
\ , 
\ \
\la{meee}
\ee
$$
\l^{-2}   = F\inv K -  C_n^2 =
1 + (Q + \td Q)r\inv + (Q\td Q - q^2_n -p^2_n) r^{-2}
- {\textstyle  {4\ov 3}} p^2_n Q r^{-3} - {\third } p^2 Q^2 r^{-4} 
\ . $$
 For $p_n\not=0$ the background 
is more   singular than in the usual
$p_n=0$ case.
This seems to be  a general pattern:
solutions with  non-vanishing
 $\bar A_i$
 coupling have  stronger short-distance singularities
(see also below).

\subsection{5-brane  model as transverse  theory }
In what follows we shall 
consider an example  with  a curved transverse theory
represented by 
the 5-brane model \re{fiv}.
Since here 
$\sqrt G e^{-2\p}  H^{kij}= \ep^{kijl}\del_l e^{-2\p}$ 
one can put the equations \re{LLL}--\re{cdw}
into the form (repeated indices are  now contracted using flat space metric)
\be
  \del^2 F^{-1} =0 \  , \ \ 
\ \ \ \ 
 \del^2 K
-   2 f\inv   {\cal F}_{ij}(\td \A)
[{\cal F}_{ij}(C)- F\inv {\cal F}_{ij}(\AA)  ]
=0 \  ,
\la{LrL}
\ee
\be
 \del_i   (f\inv  
 [{\cal F}_{ij}(\td \A) -   {\cal F}^*_{ij}(\td \A)]) =0  \ , 
\la{pp}
\ee
\be
\del_i   (f\inv  
 [{\cal F}_{ij}(C )  +   {\cal F}^*_{ij}(C )
-2 F\inv {\cal F}_{ij}(\td \A )]) 
=0\   .     
 \la{pdw}
\ee
Note the last  two  equations 
have the  following special solution 
\be
{\cal F}_{ij}(\td \A )= {\cal F}^*_{ij}(\td \A ) \ , 
\ \ \ \   \ 
{\cal F}_{ij}(\td \A ) =  \ha F [ {\cal F}_{ij}(C )  +   {\cal F}^*_{ij}(C )] \ , 
\la{self}
\ee
for which the  equation for $K$ becomes again  the 
free Laplace equation
\be
 \del^2 K = 
    f\inv   {\cal F}_{ij}(\td \A)
[{\cal F}_{ij}(C)- {\cal F}^*_{ij}(C)  ] =0  \  .
\la{LeL}
\ee
For $F=1$ this solution 
reduces to the case of self-dual $\bar \A_i$ 
and anti-self-dual $ \A_i$ 
mentioned at the end of Section 2.

\section{$D=6$ conformal  model  with  `rotations' in transverse  planes}
Let us now solve the   equations \re{LrL}--\re{pdw}
using the  $SO(4)$ symmetric  choice for the harmonic functions $f$ and $F\inv$
\be
f= 1 + Pr^{-2}\ , \ \ \ \ \  F\inv = 1 + Q r^{-2}\ , \ \ \ \ \ \ r^2=x^ix^i \ , 
\la{ppip}
\ee
and the following 
 ansatz for $\A_i, \bar \A_i$ 
 with 
rotational symmetry in  the two orthogonal planes
\be
 \A_i dx^i = h_1(r) \sin^2\t d\vp + h_2 (r) \cos^2\t d\psi\ 
 , \ \ \ 
\bar \A_i dx^i =\bar  h_1 (r) \sin^2\t d\vp + 
\bar h_2 (r) \cos^2\t d\psi\   . 
\la{ansa}
\ee
Here the
 four spatial  transverse  coordinates  $x^i$ are chosen as 
$ x^1+ix^2 = r \sin \theta e^{ i \vp },$  
\  
$ x^3 + ix^4=  r \cos \theta  e^{ i \psi },$
so that the 5-brane model Lagrangian
 $L_\bot$ in \re{fiv} takes the form 
\be
 L_\bot = f(r) [
 \del r \bd r + r^2 (\del \t \bd \t  +
 \sin^2\theta \del \vp \bd \vp   + \cos^2\t \del \psi \bd \psi)]
\la{ccr}
\ee
$$ + \  \ha P \cos 2 \t (\del \vp \bd  \psi - 
\bd \vp \del \psi)  +  
\ha {\cal R} \ln f(r) \   . 
$$
It then follows from \re{LrL} that  in general  
$K$ should depend  also on $\t$   
\be
K(r,\t) = K_1 (r) + K_2(r ) \cos 2\t  \ . 
\la{kkk}
\ee
For the gauge potential of the form in \re{ansa} one finds the following
non-vanishing components of the field strengths 
(prime denotes the derivative over $r$ 
and ${\cal F}^{ij}_\pm (\A )\equiv {\cal F}^{ij}(\A )  \pm    {\cal F}^{*ij}(\A)$) 
\be
{\cal F}^{r\vp }_\pm  = r^{-3}(r h_1' \pm 2h_2), \ \ \ \ 
{\cal F}^{r\psi }_\pm  = r^{-3}(r h_2' \pm 2h_1),  
\la{fffa}
\ee
$$
{\cal F}^{\t\vp }_\pm  = 
r^{-4}\cot \t (\pm r h_2' +  2h_1), \ \ \ \ 
{\cal F}^{\t\psi }_\pm  = - r^{-4}\tan \t (\pm r h_1' +  2h_2). 
$$
Let us first consider the special case of $\bar \A_i=0$
discussed in \ci{TT}.
Then the general solution of  the equation  \re{pdw}
for $C_i$ (i.e. for $\A_i$)  is found to be 
($k_a, n_a=\const$)
\be
 h_{+1}\equiv h_1 + \bar h_1  
=  \ha(k_1 + k_2)r^2 + k_1 P + \ha (n_1+n_2) r^{-2}
+ {\textstyle {1\ov 3} } n_2 P r^{-4}  \ , 
\la{hhhy}
\ee
$$
 h_{+2}\equiv h_2 + \bar h_2  =  \ha(k_1 - k_2)r^2 + k_1 P + \ha (n_1- n_2) r^{-2} -  {\textstyle {1\ov 3} } n_2 P r^{-4} \ .
$$
The resulting \sm \re{lrg}
is regular at $r\to 0$ only 
if there is no $r^{-4}$ pole, i.e. if $n_2=0$. 
Moreover, the background is asymptotically flat ($r\to \infty$)  if $k_1=k_2=0$. The resulting  special solution
\be
 \A_i dx^i = n_1 r^{-2} ( \sin^2\t d\vp + \cos^2\t d\psi)\ , 
\ \ \ \ \  \bar \A_i =0 \ , 
\la{ddd}
\ee
describes \ci{TT}  the $D=5$ rotating black hole \ci{MV}
with equal angular momenta in the  two orthogonal planes
$J_\vp =  J_\psi= J=n_1 {\pi/4G_N}$.   

The general solution of the 
equation \re{pp} for $\AA_i$ is similar to \re{hhhy}
\be
\td h_1 \equiv F\bar h_{1}=  \ha(q_1 +q_2)r^2 + q_2 P + \ha (p_1+p_2) r^{-2}
+ {\textstyle {1\ov 3} } p_1 P r^{-4} \ , 
\la{hhy}
\ee
$$
\td h_2  \equiv F\bar h_{2} = \ha(q_1 -q_2)r^2 -  q_2 P + \ha (p_1-p_2) r^{-2}
+ {\textstyle {1\ov 3} } p_1 P r^{-4} \ .
$$
The  self-dual solution \re{self}
corresponds to the case of  $p_1=0,\ q_2=0$.
It is easy to see that
 (independently  of the form of the associated  
solutions for $\A_i$ and $K$) 
the   \sm  \re{lrg}
has singular short distance ($r\to 0$) region  unless $p_1=p_2=0$.
The resulting background is asymptotically flat 
only if $q_1=q_2=0$. 

We conclude that there are no 
{\it  regular}  asymptotically flat 
 solutions with $\bar \A_i\not=0$. In particular, 
there does not exist a non-singular  
`extremal' 
 generalisation 
of the special solution \re{ddd} to the case of two independent 
rotational parameters.

 There is still a non asymptotically flat 
solution with  a  regular $r\to 0$ 
region which  is a natural 2-parameter generalisation 
of the horizon region  \ci{TT} of the  solution \re{ddd}.
Relaxing the condition of asymptotic flatness, i.e. 
choosing 
\be
 \AA_i dx^i =  [\ha(q_1 +q_2)r^2 + q_2 P]\sin^2\t d\vp +
[\ha(q_1 -q_2)r^2 -  q_2 P] \cos^2\t d\psi\ , 
\la{drdd}
\ee
we find that \re{pdw}
has the following
 general  solution  which is regular at $r=0$ (cf.\re{hhhy})
\be
 h_{+1}
=  \ha(k_1 + k_2)r^2 + k_1 P + \ha n_1r^{-2}
+ fF\inv (q_1 +q_2)r^2  \ , 
\la{hthhy}
\ee
$$
 h_{+2}  =  \ha(k_1 - k_2)r^2 + k_1 P + \ha n_1 r^{-2}
+ fF\inv (q_1 -q_2)r^2  \ .
$$
Then $\A_i$ is given by \re{ansa} with 
\be
h_1=  \ha(k'_1 + k'_2)r^2 + (k'_1 P + q_1Q +q_2Q)
 + \ha n'_1r^{-2}  \ , 
\la{hehy}
\ee
$$
 h_{2}  =  \ha(k'_1 - k'_2)r^2 + (k'_1 P + q_1Q-q_2Q) + \ha n'_1 r^{-2} \ , 
$$
$$ k'_1 =  k_1 +q_1, \ \ \ k'_2 = k_2+q_2, \ \ \ 
n'_1 =  n_1 + 2QPq_1\ .  $$
The function $K$  \re{kkk} can be  determined from \re{LrL}, i.e. 
from 
\be 
r^2 (r^3  K'_1)'
= 2f\inv [ r^2 \td h'_1 h'_{+1}   +  r^2 \td h'_2 h'_{+2}
+ 8\td h_1 h_{+1}   +  8 \td h_2 h_{+2}
\la{kek}
\ee
$$ -\ F\inv(  r^2 \td h'_1 \td h'_{1}   +  r^2 \td h'_2 \td h'_{2}
+ 8\td h_1 \td h_{1}   +   8\td h_2 \td h_{2}) ] \ , 
$$
\be
(r^3  K'_2)' -8r K_2  = 
r^4 [r^{-5}(r^4 K_2)']'
=  - 2r f\inv (\td h'_1 h'_{+1}   -   \td h'_2 h'_{+2})  \ . 
\la{kk}
\ee
A particular solution 
corresponding to the special case of  $k_1=k_2=n_1=0$ is
\be
K= c_0 + \td Q r^{-2}  + F\inv f r^2 (q_1^2 + q_2^2 - 2q_1q_2 \cos 2\t)  \ .  
\la{ouo}
\ee
\section{Short-distance region: 
relation to  $SL(2,R) \times SU(2)$ WZW theory}
The $r\to 0$ limit of the $D=6$ conformal 
model
discussed in the previous section  is 
 described by the following Lagrangian 
\be
L = e^{-z} \del u \bd v + 
    [Q\inv \td Q  - {P}(q_1^2 + q_2^2 - 2q_1q_2 \cos 2\t) ] \del u  \bd u
\la{lrrg}
\ee
$$ +  \  
 2P q_1  \del u  (\sin^2 \t \bd \vp + \cos^2 \t \bd \psi)
+ 2Pq_2\bd u (\sin^2 \t \del \vp - \cos^2 \t \del \psi)
$$
$$ +  \ P [
 \four \del z \bd z + \del \t \bd \t  +
 \sin^2\theta \del \vp \bd \vp   + \cos^2\t \del \psi \bd \psi
$$
$$ + \  \ha \cos 2 \t (\del \vp \bd  \psi - 
\bd \vp \del \psi) ] \   , \ \ \ \   z\equiv   - \ln ( Q\inv r^2) \ . 
$$
The  total dilaton 
$\P = \ha \ln (Ff)$ is constant in the $r\to 0$  limit. 
This model is related to 
 the $SL(2,R) \times SU(2)$ WZW model\footnote{The standard $SL(2,R)$ WZW Lagrangian  written in the Gauss decomposition 
parametrisation can be put in the  `non-standard' 
form which appears here by the following field redefinition 
\ci{TH}
$$ e^{-2x'} \del u' \bd v' + \del x' \bd x'=
e^{-2x} \del u \bd v + \del u \bd u +     \del x \bd x , \ \ \ 
u'= \ha  e^{2 u}, \ \ \ v'= v - e^{2 u}, \ \ 
\  x'= x + u. $$
The levels of $SU(2)$ and $SL(2,R)$ WZW models are both equal to $P/\a'$.
}
\be
   L =   e^{-z} \del u \bd v + 
    Q\inv \td Q  \del u  \bd u  +  \four P  \del z \bd z 
\la{lryrg}
\ee
$$+ \  P( \del \t \bd \t  +
 \sin^2\theta \del \vp \bd \vp   + \cos^2\t \del \psi \bd \psi)
+   \ha  \cos 2 \t (\del \vp \bd  \psi - 
\bd \vp \del \psi)]  \   ,  
$$
by the formal 
 coordinate shift\foot{The  standard $SU(2)$ Euler angles ($\t',\vp',\psi'$)
are related to the angular coordinates  used above  by
$$\t= \ha \t', \ \ \ \vp= \ha (\vp' + \psi'), \   \ \ \psi=  \ha (\psi' - \vp'), \ \ \  0 \leq \t' \leq \pi, \   0 \leq \vp' \leq 2\pi,\ 
0 \leq \psi' \leq 4\pi. $$
}
\be
\vp \to \vp + (q_1 +q_2) u \ , \ \ \ \ 
\  \psi \to \psi + (q_1 -q_2) u \ .
\la{rer}
\ee
Thus  the throat region model \re{lrrg}  is locally a direct product
(but is globally non-trivial for generic values of $q_1,q_2$).
It  has broken supersymmetry \ci{TSE}
unless the twists
$q_1,q_2$ take quantised values 
(for which the resulting string model becomes  equivalent to the model with $q_1=q_2=0$). 
Note that  the positivity of the coefficient of  $\del u \bd u$ 
term in both   \re{lrrg} and 
 \re{lryrg} implies  a  bound  on $|q_1-q_2|$:
 $\ (q_1-q_2)^2 < Q\inv P\inv \td Q$.

  Though this model 
has two natural `twist' parameters,  we have found above  that  for $q_2\not=0$ 
it does not have 
an  extension to finite $r$  which is 
asymptotically flat.
It may be of interest to study related 
 non asymptotically flat models from the point of view 
 of their possible `magnetic' interpretation, given that 
the corresponding backgrounds 
 can be  transformed 
into solutions   with non-vanishing R-R fields
 which have straightforward $D$-brane interpretation.

\bigskip
Non-extremal 
5-dimensional black hole solutions
with two rotational parameters were recently constructed in 
 \ci{all,Mir}.

 \section*{Acknowledgments}
I am  grateful to M. Cveti\v c and  J. Maldacena
 for   useful
discussions. 
This work was  supported by  PPARC,   
 ECC grant SC1*-CT92-0789 and NATO  grant CGR 940870.


\end{document}